\documentclass[a4paper]{article}
\usepackage{INTERSPEECH2022}
\usepackage{CJKutf8, cite}
\usepackage{multirow}
\usepackage{makecell}
\usepackage{color}
\DeclareMathOperator*{\argmax}{argmax}
\newcommand\blfootnote[1]{%
\begingroup
\renewcommand\thefootnote{}\footnote{#1}%
\addtocounter{footnote}{-1}%
\endgroup
}
\title{Internal Language Model Estimation based Language Model Fusion for Cross-Domain Code-Switching Speech Recognition}
\name{ Yizhou Peng$^{1,4\dag}$, Yufei Liu$^{2,4\dag}$, Jicheng Zhang$^{1,4\dag}$, \\ Haihua Xu$^3$, Yi He$^3$, Hao Huang$^{1,5*}$, Eng Siong Chng$^4$}

\address{
  $^1$SISE, Xinjiang University, Urumqi, China\\
  $^2$South China University of Technology, Guangzhou, China\\
  $^3$AI Lab ByteDance\\
  $^4$SCSE, Nanyang Technological University, Singapore\\
  $^5$Xinjiang Provincial Key Laboratory of Multi-lingual Information Technology, Urumqi, China}
\email{}

\begin{document}

\maketitle
\begin{abstract}
 Internal Language Model Estimation (ILME) based language model (LM) fusion has been shown significantly improved recognition results over conventional shallow fusion in both intra-domain and cross-domain speech recognition tasks. In this paper, we attempt to apply our ILME method to cross-domain code-switching speech recognition (CSSR) work. Specifically, our curiosity comes from several aspects. First, we are curious about how effective  the ILME-based LM fusion is for both intra-domain and  cross-domain  CSSR tasks. We verify this with or without merging two code-switching domains. More importantly, we train an end-to-end (E2E) speech recognition model by means of merging two monolingual data sets and observe the efficacy of the proposed ILME-based LM fusion for CSSR. Experimental results on SEAME that is from Southeast Asian and another Chinese Mainland CS data set demonstrate the effectiveness of the proposed ILME-based LM fusion method.
 % obtaining up to 8\% and 37.3\% relatively improvements respectively.
\end{abstract}

\blfootnote{
    \quad  $^*$correspondence author.
}
\blfootnote{
\quad This work is done when $\dag$ are working  as interns at SCSE, MICL Lab, NTU, Singapore.
}
\section{Introduction}
Code-switching (CS) refers to a discourse containing more than one language in terms of  written or spoken form. On utterance level, one can further classify CS into two categories. One is there is different language alternation within an utterance, normally called as intra-sentential code-switching, here intra-CS for brevity. Another is language alternation occurs across neighboring utterances, here called as inter-CS.
CS ASR \cite{guo18b_interspeech,zeng19_interspeech,shi2020asru,shan2019ICASSP,li2019ICASSP,peng2021minimum,khassanov2019constrained,ma2019APSIPA,zhang20r_interspeech} is more challenging when it is compared with monolingual-based speech recognition  due to data sparsity issue. This is particularly true for the intra-CS case.
In such a case, the code-switching is not only culture-dependent it is also personal peculiarity-dependent. As a result, how code-switching occurs is full of uncertainty, and such an uncertainty results in data sparsity, and hence poor recognition results. 

To achieve an improved code-switching speech recognition performance, the first strategy one can easily think of is to employ data augmentation approaches~\cite{ko15_interspeech,Du-SLT2021,ylmaz18_interspeech}.  For instance, one can employ off-the-shelf text-to-speech system to generate more audio-text paired CS data to train ASR models~\cite{Du-SLT2021}. However, the difficulty lies in a text-to-speech system capable of synthesizing spontaneous code-switching data itself doesn't easily come by. 
Likewise, another easier avenue is to generate more CS text to boost language modeling performance. \cite{2020CS-MT} employs Machine Translation (MT) and  word/phrase substitution methods to generate more CS text. The difficulty is how to get more paired text data to yield a sensible MT model.  Recently, \cite{li20l_interspeech} proposed to use CycleGan~\cite{zhu2017ICCV} to produce CS text data without paired text data to train the sequence-to-sequence model.  Similarly aiming to alleviate CS data sparsity issue, \cite{liu2021ICASSP} translates CS recognition results to monolingual text to score.

Except for explicit data augmentation method addressing data scarcity issue, one can aim to approach the problem from model optimization perspective. \cite{chuang2020ICASSP} proposed to employ multilingual data to train CS E2E ASR model by penalizing  the divergence of ASR output distributions of different languages, while \cite{ylmaz18_interspeech,chang19_interspeech} employed the similar idea to train neural network CS language model with monolingual text data. Also through model optimization efforts, \cite{li2019ICASSP} proposed to employ LID classification results to fine-tune the output of a CTC-based E2E CS ASR score, yielding improved performance on intra-CS recognition. More recently, \cite{2022_context_confusion} proposed an  language-dependent self-attention mechanism that is  to reduce cross-lingual context confusion and achieved improved results. The underlying motivation of \cite{yan2021joint} is to fully take advantage of monolingual data and make what is learned from monolingual data be transferable to CS ASR models. Similar idea is also reflected in \cite{zhou20b_interspeech}, where two monolingual encoders are employed to pretrain the ASR model, and less CS data is employed to  fine-tune the pretrained model, attaining improved CS ASR results.

% \textcolor{red}{ILME methods are motivated by the density ratio method~\cite{} and hybrid autoregressive transducer(HAT)~\cite{} for more efficiently integrating external language models~\cite{}, where language model integration~\cite{} is a promising method for adapting or strengthening the performance of End-to-End models, for either intra-domain or cross-domain speech recognition.}
In this paper, we are interested in cross-domain CS ASR instead. Concretely, we improve cross-domain CS ASR performance by internal language model estimation (ILME)~\cite{meng2021internal-estimate,meng2021internal-train,variani2020hybrid,mcdermott2019density,zeyer2021librispeech,sugiyama2012density} based language model (LM) fusion~\cite{sriram2017cold,shan2019component,zeineldeen2021investigating,zhao2019shallow,le2021deep} method. 
This is inspired from the prior works where ILME method has significantly improved both intra-domain and cross-domain ASR performance in diversified monolingual ASR tasks. As mentioned, cross-domain CS ASR is much more challenging in contrast with monolingual ASR, since CS  is not only culture-dependent it is also personal peculiarity-dependent, especially for the intra-CS case. For instance Chinese mainland English-Mandarin CS is remarkably different from Southeast Asian CS in terms of accent, as well as style. Chinese mainland CS is a  mandarin-predominant CS with single English content word in utterance in most cases, while in Singapore of Southeast Asia, English and Mandarin word ratio is more diversified in daily conversation. Besides, there are frequent connective words appearing in utterance, such as \begin{CJK*}{UTF8}{gbsn} ``Then 我就 去 canteen 吃饭了 (Then I went to canteen for meal)"\end{CJK*}. 
% This is inspired from our prior work where our proposed ILME method~\cite{ILME-Liuyufei} has significantly improved both intra-domain and cross-domain ASR performance in diversified monolingual ASR tasks. As mentioned, cross-domain CS ASR is much more challenging in contrast with monolingual ASR, since CS  is not only culture-dependent it is also personal perculiarity-dependent, especially for the intra-CS case. For instance Chinese mainland English-Madarin CS is remarkably different from Southeast Asian CS in terms of accent, as well as style. Chinese mainland CS is a  mandarin-predominant CS with single English content word in utterance in most cases, while in Singapore of Southeast Asia, English and Madarin word ratio is more diversified in daily conversation. Besides, there are frequent connective words appearing in utterance, such as \begin{CJK*}{UTF8}{gbsn} ``Then 我就 去 canteen 吃饭了 (Then I went to canteen for meal)"\end{CJK*}. 

Specifically, in this paper we examine the efficacy of the proposed ILME-based LM fusion method~\cite{ILME-Liuyufei} for cross-domain CS ASR performance between SEAME~\cite{lyu2010seame} and a mainland Chinese-dominated CS data set that is release by \texttt{DataTang}  for ASRU 2019 English-Mandarin CS ASR Challenge~\cite{shi2020asru}.  Our contributions are mainly reflected in the following aspects:
\begin{enumerate}

    \item To the best of our knowledge, this is the first pioneering work showing the efficacy of the ILME-based LM fusion for CS ASR, in terms of both intra-domain and cross-domain scenarios under E2E ASR framework.
    \item More importantly, we attempt to employ monolingual data to train multilingual ASR for CS recognition, employing ILME-based LM fusion, leading to significant performance improvement compared with conventional shallow fusion.
    \item We elaborate how to perform ILM estimation using Transformer decoder.
\end{enumerate}

% \section{related works}

\section{Proposed Method}~\label{sec:proposed}
\vspace{-8mm}
\subsection{ILME-based LM fusion}~\label{sub:ilme}
In this paper, we use Conformer for E2E ASR modeling. With the trained conformer, the ASR inference process can be described as follows:
\begin{equation}
  \hat{Y} = \argmax_{Y}\log P(Y|X)\label{eq:asr-infer}
\end{equation}
where $X$ is input acoustic features, and $Y$ is inferred word sequence, while $P(Y|X)$ is a joint posterior probability.

Give external text data that is employed to train a LM, the E2E ASR inference is formulated as:
\begin{equation}
  \hat{Y} = \argmax_{Y}(\log P(Y|X) + \lambda \log P_{\text{LM}}(Y) )\label{eq:e2e-asr-lm-infer}
\end{equation}
where $P_{\text{LM}}(Y)$ is the joint probability estimated from external LM. What Eq. \ref{eq:e2e-asr-lm-infer} states is not harmony with Bayes probability theory for conventional hybrid DNN-HMM ASR, where it states:
\begin{equation}
   \hat{Y} = \argmax_{Y}(\log P(X|Y) + \lambda \log P_{\text{LM}}(Y) )\label{eq:hybrid-asr-infer}
\end{equation}
To realize LM fusion as in conventional DNN-HMM ASR under E2E ASR framework,  we approximate $\log P(Y|X)$ in Eq.~\ref{eq:e2e-asr-lm-infer} as $\log P(Y|X) := \log P(X|Y) + \log P(Y)$ using Bayes posterior probability. Replacing term $\log P(X|Y)$ in Eq.~\ref{eq:hybrid-asr-infer}, we have Eq.~\ref{eq:hybrid-asr-infer01}:
\begin{align}
  \hat{Y} &= \argmax_{Y}(\log P(Y|X) - \log P(Y) \nonumber \\
          & + \lambda\log P_{\text{LM}}(Y) )\label{eq:hybrid-asr-infer01}
\end{align}
where $P(Y)$ is estimated with training transcripts, as a result we called as internal LM (ILM), i.e., $P_{\text{ILM}}(Y)$. Besides, we normally need to rescale it, and  Eq.~\ref{eq:hybrid-asr-infer01} is turned into:
% However using Bayes posterior theory,  Eq.~\ref{eq:asr-lm-infer} can be further written as:
% \begin{equation}
%   \hat{Y} = \argmax_{Y}(\log p(X|Y) + \log p_{\text{ILM}}(Y) + \lambda \log p_{\text{LM}}(Y) )\label{eq:asr-ilm-decompose-infer}
% \end{equation}
% where $p_{\text{ILM}}(Y)$ is implicitly estimated with E2E ASR models that are learned from training transcripts. Using Eq.~\ref{eq:asr-ilm-decompose-infer}  for LM fusion is suboptimal, especially when the ASR is applied to cross-domain speech recognition. Consequently, if we want to realize E2E-based speech recognition with LM fusion that is consistent with Bayes posterior theory, we should practice as follows:
\begin{align}
  \hat{Y} &= \argmax_{Y}(\log P(Y|X) - \lambda_{\text{ILM}}\log P_{\text{ILM}}(Y) \nonumber \\
          & + \lambda\log P_{\text{LM}}(Y) )\label{eq:e2e-asr-ilm-infer}
\end{align}
where $\lambda_{\text{ILM}}$ is ILM scaling factor. Now, we can apply Eq.~\ref{eq:e2e-asr-ilm-infer} for E2E ASR LM fusion that is in harmony with Bayes theory for ASR, and the overall focus is on how to estimate the internal LM score $\log P_{\text{ILM}}(Y)$ with the existing ASR framework.

\subsection{Transformer-decoder-based ILME}~\label{sub:transformer-decoder}
\begin{figure} [th]
    \centering
     \includegraphics[width=0.8\linewidth]{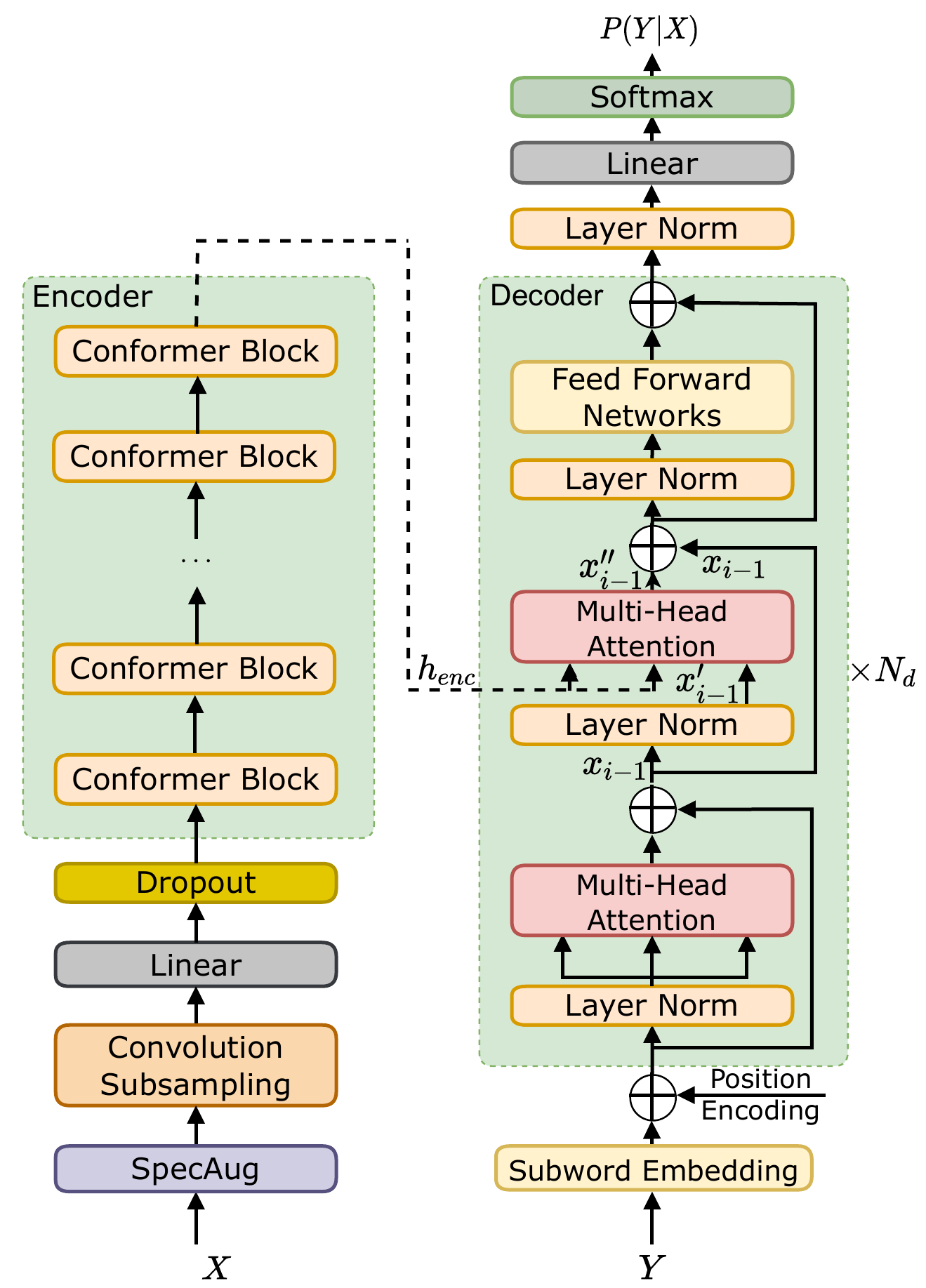}
    \caption{Illustration of Conformer architecture for corresponding ILME (One only needs to focus on cross-attention module)}
    \label{fig:conformer-decoder}
\end{figure}
To realize Transformer-decoder-based ILME, we need to revisit cross-attention modules in each decoder block as indicated in Figure~\ref{fig:conformer-decoder}.

For each cross-attention, we have the following operations:
\begin{align}
    &x_{i-1}^{\prime} = \text{Layernorm}(x_{i-1}) \label{eq:layernorm} \\
    &x_{i-1}^{\prime\prime} = \text{MHCA}(x^{\prime}_{i-1}, h_{\text{enc}}) \label{eq:ca} \\
    &x_i = x_{i-1}^{\prime\prime}  + x_{i-1} \label{eq:res}
\end{align}
where Layernorm and MHCA refers to layer normalization and multihead cross-attention operations respectively.
For ILME, we need to make the decoder act as an LM model, and the parameters of such an LM model should be maximally shared with existing decoder. 
% that is, we want $x^{\prime\prime}_{i-1}$ in Eq.~\ref{eq:ca} to be estimated ignoring encoder output $h_{\text{enc}}$.  
Following what is proposed in \cite{ILME-Liuyufei}, we employ two methods to estimate ILM.  We name the first method as One-time Context-vector Learning (OTCL). Here, the context-vector refers to $x^{\prime\prime}_{i-1}$ in Eq.~\ref{eq:ca}. By 
OTCL-based IMLE, it is estimated as follows:
\begin{equation}
    x_{i-1}^{\prime\prime} = b \label{eq:otcl}
\end{equation}
where b is a learnable bias in Eq.~\ref{eq:res}. For simplicity, it is shared across different decoder blocks and it is learned with overall training transcripts by fixing the other learned parameters in decoder.

The obvious limitation of Eq.~\ref{eq:otcl} is that  the bias cannot remember the history context, besides, once it is learned it cannot change. As a result, \cite{ILME-Liuyufei} proposes another IMLE method, denoted as label synchronous context-vector  learning (LSCL). Specifically, we introduce a lightweight Feed Forward Network (FFN), realizing:
\begin{equation}
       x_{i-1}^{\prime\prime} = \text{FFN}(x^{\prime}_{i-1})~\label{eq:lscl}
\end{equation}
Likewise, the FFN is shared by different decoder blocks.

\section{Experiments}

\subsection{Data}

We select two Mandarin-English code-switching ASR data sets  for cross-domain CS ASR performance study. One is SEAME~\cite{lyu2010seame}, a conversational Mandarin-English corpus from Southeast Asia, i.e., Malaysia and Singapore, another is a Mandarin-English CS data set from Chinese Mainland, released by \texttt{Datatang} for an open Mandarin-English CS ASR challenge in ASR 2019~\cite{shi2020asru}. For brevity, we name it as ASRU in what follows.
Though both data sets are Mandarin-English CS, they are hugely different. They are from different areas that means CS influenced with different cultural background.
More importantly, SEAME is conversational speech, while ASRU is reading speech, and hence much simpler. Table~\ref{tab:speech-data} reports overall speech data distributions in details.
\begin{table}[htp]
 \centering
\caption{Overall Speech Data distribution for both ASR model training and testing, while ``Man/Eng" is the ratio of Mandarin Characters to English Words for specific data sets.}
 \label{tab:speech-data}
\begin{tabular}{cccc}
 \toprule
  Corpus & Subset & Duration(Hrs) & Man/Eng %man/eng 
  \\
   \midrule
   \multirow{3}*{SEAME} & Train & 93.6 & 0.83/0.17\\
   & Dev\textsubscript{Man} & 7.5 & 0.91/0.09\\
   & Dev\textsubscript{Sge} & 3.9 & 0.55/0.45\\
\midrule
    \multirow{4}*{ASRU} & Train & 193.0 & 0.97/0.03 \\
    & Dev1 & 20.4 & 0.97/0.03 \\
    & Dev2 & 21.3 & 0.98/0.02\\
    & Test & 20.6 & 0.98/0.02\\
\midrule
    \multirow{2}*{Multi-Lingual} & Mandarin & 100.6 & \multirow{2}*{0.53/0.47} \\
    & English & 100.5 & \\
    \bottomrule
\end{tabular}
\end{table}
Since we are not only interested in ILME-based LM fusion for cross-domain CS ASR, we are also deeply concerned with to what extend IMLE-based LM helps CS ASR for E2E ASR models that are trained with two monolingual data sets, i.e., English and Mandarin data sets here. To this end, we make a new ``Multi-lingual"  data set by merging two monolingual data sets in Table~\ref{tab:speech-data}.

% Except for code-switching scenario, 
% \textcolor{red}{we are also interested in if multilingual ASR model is capable of recognizing CS speech,} 
% then we merge two monolingual data to realize a Mandarin-English mixed corpus named Multi-Lingual to train a ASR model.

% To study cross-lingual code-switching ASR, we focus on two CS speech corpora, which are South-East-Asia Mandarin-English and Chinese mainland Mandarin-English Code-Switching speech from ASRU 2019 English-Mandarin CS ASR Challenge. For simplicity, we name them as SEAME and ASRU respectively. SEAME data is recorded under clean conversational environment, and ASRU data is recorded under clean environment in reading style.

% then we reorganized a Multi-Lingual corpus containing quantification balanced Mandarin and English data. 

To be specific, ``Man/Eng" in Table~\ref{tab:speech-data} refers to average ratios of Mandarin characters to English words. From Table~\ref{tab:speech-data}, we can see that ASRU CS data is much more predominated with Mandarin. Actually, majority of sentences are dotted with only a single English word. Besides, our SEAME data set definition follows~\cite{peng2021minimum}, while ASRU also follows official data set definition~\cite{shi2020asru}. Finally, For ``Multilingual " corpus, we get majority of English speech data from IMDA corpora~\cite{koh19_interspeech_imda}, while Mandarin data is from extended \texttt{DataTang} release~\cite{shi2020asru}.

% Table~\ref{tab:speech-data} illustrates the information of speech data used in our ASR models. 
% \textcolor{red}{The Man/Eng ratio is calculated in word level for English and character level for Mandarin.
%  For SEAME data, we follow the same data definition as~\cite{peng2021minimum}. There are about 105 hours of speech in total, where Dev\textsubscript{Man} and Dev\textsubscript{Sge} are two evaluation subsets.
%  % from non-overlapping speakers of Train subset. 
%  Dev\textsubscript{Man} is dominated with Mandarin while Dev\textsubscript{Sge} is dominated with Singapore English.}
%  \textcolor{red}{For ASRU data, training set is around 200 hours, and three subsets of evaluation named Dev1, Dev2 and Test contain 20 hours speech of each, 
%  with similar Man/Eng ratio as training set.
%  The Multi-Lingual corpus contains 100 hours of Mandarin and English speech respectively, where Mandarin is dominated with Mainland Mandarin and English is from Singapore area.}

Language modeling is decisive in this paper. To realize a comprehensive study of how ILME-based LM fusion works under different CS scenarios,  we define 6 data sets to build diversified LMs as indicated in Table~\ref{tab:text-data}, where  ``SEAME-CS" and ``ASRU-CS" are corresponding training transcripts, while ``SEAME-CS-extra" includes more IMDA~\cite{koh19_interspeech_imda} CS data and ``ASRU-CS-extra" includes  500 hours of Mandarin transcripts provided by \texttt{DataTang} as well as 960 hours of transcripts from Librispeech, and finally, the ``SEAME-ASRU-*" refers to the text data by corresponding data combinations.
 
% \textcolor{red}{For language modeling, we employ six different subsets of text, belonging to three domains which are SEAME, ASRU and combined SEAME-ASRU. Table~\ref{tab:text-data} shows the six language models namely LM1 to LM6, and corresponding training text.}
 
\begin{table}[htp]
 \centering
% \caption{LM Training Text Data. SEAME-CS and ASRU-CS are training transcriptions of SEAME and ASRU data respectively. SEAME-CS-Extra is the combination of SEAME transcriptions and conversational text from south-east Asia areas. ASRU-CS-Extra is ASRU training transcriptions combined with Librispeech960 training transcriptions and extra 500h of Mandarin text from ASRU data. SEAME-ASRU-CS is the combination of SEAME-CS and ASRU-CS. SEAME-ASRU-CS-Extra refers to SEAME-CS-Extra combined with ASRU-CS-Extra. }
\caption{Overall text data for external LM training.}
 \label{tab:text-data}
\begin{tabular}{clcc}
 \toprule
  ID & \makecell[l]{Training Text} & \#utts & \#words \\
   \midrule
   LM1 & SEAME-CS & 284K & 4.8M \\
   LM2 & SEAME-CS-Extra & 5.6M & 65.9M\\
   LM3 & ASRU-CS & 180K & 1.8M \\
   LM4 & ASRU-CS-Extra & 1.0M & 17.0M \\
   LM5 & SEAME-ASRU-CS & 465K & 6.6M \\
   LM6 & SEAME-ASRU-CS-Extra & 6.7M & 82.9M\\
    \bottomrule
\end{tabular}
\end{table}

\subsection{Models}~\label{sub:models}
The ASR models are built with Conformer~\cite{2020Conformer} and LMs are built with Transformer using Espnet toolkit~\cite{watanabe2018espnet}. For ASR models, Conformer encoder is configured with 12-layer, and the decoder is configured with 6-layer with 4-head attentions. The attention dimension is 256. The models are trained with CTC/Attention criteria, of which the weighting factor for CTC is 0.3. The input acoustic features are 83-dim that is composed of filter-bank and pitch features. For external LMs, they are configured with 16-layer Transformer decoder with 8-head 512-dim attentions. The output of both models are BPE with 3000 English word pieces and 3930 Chinese characters. For the LSCL-based ILME, the lightweight FFN has two full connected layers and each has 128 units.

For training, we employ Noam optimizer~\cite{vaswani2017attention} with dropout rate of 0.1 to all of our models, and the final models are obtained by averaging the  parameters of the 10 models that have the best accuracy on validation set. For inference, the weighting factor of CTC is increased to 0.4, and all results are reported with Mix Error Rate (MER), that is, character for Chinese while word for English.
% \textcolor{red}{
% Our ASR models are conducted with Conformer and LMs are conducted with Transformer on Espnet toolkit~\cite{watanabe2018espnet}. For ASR models, Conformer is configured with 12-layer encoder, and 6-layer decoder with 4-head attention, the attention dimension is 256. The models are jointly trained with CTC/Attention criterion, of which the weighting factor for CTC is 0.3. The input acoustic features are 83-dim Fbank-Pitch. 
% % with 80-dim Filter Banks and 3-dim Pitch.
% For LMs, they are configured with 16-layer decoder with 8-head attention, where attention dimension is 512.
% The output of the models is BPE set consisting of 3000 English subwords and 3920 Chinese characters.
% For LSCL method based ILME, the lightweight FFN has two layers and each has 128 units.
% }

% \textcolor{red}{We apply Noam optimizer~\cite{vaswani2017attention} with dropout rate of 0.1 to all of our models, and the final models are obtained by averaging the model parameters of the best 10 models with the highest valid accuracy.
% In inference period, the weighting factor of CTC is fixed to 0.4. All the results are reported in MER(Mix Error Rate)\% evaluation matrix.}

\section{Results} 
% \textcolor{red}{
% Table~\ref{tab:ppl} indicates the perplexity between each LM and evaluation sets. Compared between LM1\&LM2 and LM3\&LM4, we found that LMs in the specific domain perform much lower PPL for intra-domain text, compared with cross-domain ones. This shows there is a mismatch between SEAME and ASRU domain in text respect.  
% % For LMs in SEAME domain, which are LM1 and LM2, produce much lower PPL for SEAME test sets compared with ASRU. In contrast, for LMs in ASRU domain, produce reversed results compared with LMs in SEAME domain.  % But LM3 get high PPL among all testsets.
% LM5 and LM6 perform low PPL on all text sets as they are combined with the domain of SEAME and ASRU.}
% Among all the perplexity results from table~\ref{tab:ppl}, we found that there is a serious mismatch between SEAME and ASRU domain in language respect.
Table~\ref{tab:ppl} presents our perplexity results on overall test sets using all our LMs in Table~\ref{tab:text-data}.
We observe from Table~\ref{tab:ppl} that cross-domain perplexity is quite high, implying significant difference between two CS data sets. Besides, LM5\&LM6 by merging two CS text data leads to remarkable perplexity drop for two CS test sets.
It is noteworthy it gets significantly better perplexity when more English and Mandarin data is merged with ASRU transcript data (see LM4 versus LM3).
\begin{table}[htp]
 \centering
\caption{Perplexity for each evaluation sets, based on subwords}
 \label{tab:ppl}
\begin{tabular}{ccc|ccc}
 \toprule
  \multirow{2}*{ID} & \multicolumn{2}{c|}{SEAME} & \multicolumn{3}{c}{ASRU} \\
  & Dev\textsubscript{Man} & Dev\textsubscript{Sge} & Dev1 & Dev2 & Test \\
   \midrule
   LM1 & 52 & 78 & 190 & 256 & 249 \\
   LM2 & 47 & 57 & 119 & 145 & 142 \\
   LM3 & 388 & 28128 & 31 & 399 & 4607 \\ % 科学计数法
   LM4 & 704 & 686 & 38 & 236 & 228 \\
   LM5 & 82 & 108 & 43 & 172 & 168 \\
   LM6 & 49 & 58 & 44 & 105 & 102 \\
    \bottomrule
\end{tabular}
\end{table}

% \textcolor{red}{
% Table~\ref{tab:baseline} shows all results for baseline Code-Switching ASR systems. First, for the SEAME systems, S1 performs well on evaluation sets from SEAME, but poor performance showed on ASRU evaluation sets. Vise versa, for ASRU systems, A1 produces much better results on ASRU evaluation sets, but can barely recognize the sentences from SEAME, especially for Dev\textsubscript{Sge} which contains more English words. While we combine SEAME and ASRU data, we surprisingly find that C1 consistently outperforms both of A1 and S1 on all evaluation sets which suggests that even huge mismatch exists between different domains, there is still shared beneficial information.} 

\begin{table}[htp]
 \centering
% \caption{Recognition Results of Baseline ASR models}
\caption{MERs(\%) with ILME-based LM fusion for CS cross-domain ASR.}
 \label{tab:cs-cd-asr}
\begin{tabular}{p{10pt}lp{20pt}p{20pt}|p{10pt}p{10pt}p{10pt}}
 \toprule
 \multirow{2}{*}{ID} & \multirow{2}*{\makecell[c]{Models}} & \multicolumn{2}{c|}{SEAME} & \multicolumn{3}{c}{ASRU}  \\
  & & Dev\textsubscript{Man} & Dev\textsubscript{Sge} & Dev1 & Dev2 & Test \\
 \midrule
 \multicolumn{7}{c}{\textbf{SEAME} (LM weight=0.4)} \\
 S1 & Baseline & 16.4 & 23.2 & 63.6 & 58.9 & 58.4 \\
 S2 &  +SF-LM1 & 15.9 & 22.4 & 63.5 & 59.9 & 59.0 \\
 S3 & +SF-LM2 & 15.7 & 21.9 & 62.3 & 58.3 &57.4 \\
 S4 & +SF-LM4 & 18.4 & 25.3 & 56.6 & 56.8 &56.1 \\
 S5 & +OTCL + SF & 15.5 & 22.1 & 51.8 & 56.2 & 54.8\\
 S6 & +LSCL + SF & 15.4 & 21.7 & 49.5 & 54.3 & 54.7\\
 \midrule
 \multicolumn{7}{c}{\textbf{ASRU} (LM weight=0.1)} \\
 A1 & Baseline & 67.9 & 86.9 & 8.6 & 14.0 &13.2\\
 A2 & +SF-LM3 & 67.6 & 86.7 & 7.9 & 13.9 & 13.1\\
 A3 & +SF-LM4 & 67.3 & 86.4 & 8.0 & 13.7 & 12.9\\
 A4 &  +SF-LM2 & 65.7 & 86.2 & 8.3 & 13.4 & 12.5\\
%  \quad + SF-LM6 & 66.1 & 86.2 & 12.8\\
 A5 & +OTCL + SF & 64.5 & 86.1 & 8.0 & 14.0 & 13.3 \\
 A6 & +LSCL + SF & 65.3 & 86.0 & 7.9 & 13.8 & 13.2 \\
 \midrule
 \multicolumn{7}{c}{\textbf{SEAME + ASRU} (LM weight=0.1)} \\
 C1 & Baseline & 15.4 & 22.2 & 8.3 & 12.7 & 12.0 \\
 C2 &  +SF-LM5 & 15.1 & 21.8 & 7.9 & 12.5 & 11.9 \\
 C3 & +SF-LM6 & 15.0 & 21.6 & 7.8 & 12.1 & 11.4 \\
 C4 & +OTCL + SF & 14.8 & 21.3 & 7.8 & 12.0 & 11.5 \\
 C5 &  +LSCL + SF & 14.5 & 20.9 & 7.4 & 11.5 & 10.9 \\
    \bottomrule
\end{tabular}
\end{table}

Table~\ref{tab:cs-cd-asr} reports our overall CS cross-domain ASR results. 
We have trained 3 ASR models using SEAME, ASRU, as well as SEAME + ASRU speech data respectively. 
For all IMLE-based LM fusion methods, we have the following configurations to mention.
In S5\&S6 of the SEAME ASR group, we apply LM2 on SEAME test sets, and apply LM4 to ASRU test sets.
The same LM2 and LM4 are also applied for A5\&A6 of the ASRU ASR group respectively.
For C4\&C5 of the SEAME+ASRU ASR group, LM6 are applied for both.

For experiments with SEAME ASR, we observe the proposed LSCL-based IMLE fusion method makes consistent better recognition results  compared with shallow fusion methods
in both intra-domain (for SEAME) and cross-domain (for ASRU) CS ASR tasks.
The MER improvement in intra-domain is minor. It can be seen when S6 is compared with S3, where
the MER improvement is from 15.7\% to 15.4\% for Dev\textsubscript{Man}, and 
21.9\% to 21.7\% for Dev\textsubscript{Sge}. This is understandable since our external LM text data is rather limited and majority are monolingual data.
The MER improvement for the cross-domain is slightly better as is seen by comparing S6 with S4. 
% The MER drops are from 56.6\% to 49.5\% on Dev1, from 56.8\% to 54.3\% on Dev2, and 56.1\% to 54.7\% on Test respectively. We believe if given more target-domain relevant text data, ILME can be more beneficial.
% in terms of performance improvement.

For experiments with ASRU ASR, table~\ref{tab:cs-cd-asr} shows the proposed method also brings  performance improvement over the shallow fusion method in cross-domain experiments though very minor. However we haven't observed any improvement over shallow fusion in ASRU intra-domain case.

% For the poor performance in cross-domain experiments, we think it is relevant with poor ASR models for cross-domain recognition. 
% As is shown in Table~\ref{tab:cs-cd-asr}, ASRU ASR system has 86.9\% MER on the Dev\textsubscript{Sge} test set. This clearly reveals the main difference of two CS corpora mainly arise from English difference.
% % Besides, we don't think Librispeech English data is close to  English data in SEAME.
% While for the poor performance in intra-domain results, we think it is due to weak language models since very limited text data is employed (See Table~\ref{tab:text-data}). As a result ILME fails to give improvement.

Finally, Table~\ref{tab:cs-cd-asr} also reports various recognition results from ASR models trained with SEAME and ASRU combined data. Surprisingly, we get best baseline ASR results on overall test sets of two corpora. More importantly, both IMLE-based fusion methods are effective for MER reduction compared with shallow fusion method.

In Table~\ref{tab:cs-cd-asr}, we have demonstrated the efficacy of the proposed methods in terms of cross-domain CS ASR, and all ASR models are trained with CS data. However, CS data is hard to access in reality. Now we want to examine how an ASR model that is trained with combined monolingual data performs for CS ASR, particularly with the help of ILME-based LM fusion method. Table~\ref{tab:multilingual-result} reports the MER results with the multilingual ASR models using ILME-based LM fusion method. We notice that in all cases the proposed method makes significant MER reduction over the shallow fusion method. This is especially remarkable in the ASRU case, and we get up to 32.06\% WERR.

\begin{table}[htp]
 \centering
\caption{MERs(\%) of Multilingual ASR models with ILME-based LM fusion method.}
 \label{tab:multilingual-result}
\begin{tabular}{lcc|ccc}
 \toprule
 \multirow{2}*{\makecell[c]{Models}} & \multicolumn{2}{c|}{SEAME} & \multicolumn{3}{c}{ASRU} \\
 & Dev\textsubscript{Man} & Dev\textsubscript{Sge} & Dev1 & Dev2 & Test \\
 \midrule
 Baseline& 33.0 & 38.4 & 65.2 & 64.7 & 63.9  \\
 \quad + SF-LM2 & 31.6 & 37.2 & - & - & -\\
 \quad \quad + LSCL & \textbf{30.1} & \textbf{35.8} & - & - & - \\
 \quad + SF-LM4 & - & - & 60.2 & 62.3 & 61.6 \\
 \quad \quad + LSCL & - & - & \textbf{40.9} & 49.7 & 48.6 \\
 \quad + SF-LM5 & 32.1 & 37.6 & 60.2 & 61.7 & 61.0 \\
 \quad \quad + LSCL & 31.3 & 37.1 & 42.9 & 48.1 & 47.1 \\
 \quad + SF-LM6 & 31.9 & 37.4 & 61.6 & 62.8 & 62.1 \\
 \quad \quad + LSCL & 30.4 & 36.1 & 42.9 & \textbf{46.5} & \textbf{45.6} \\
     \bottomrule
\end{tabular}
\end{table}

\section{Analysis}
\begin{figure}[th]
  \centering
    \begin{minipage}[t]{.497\linewidth}
      \includegraphics[width=1.0\linewidth]{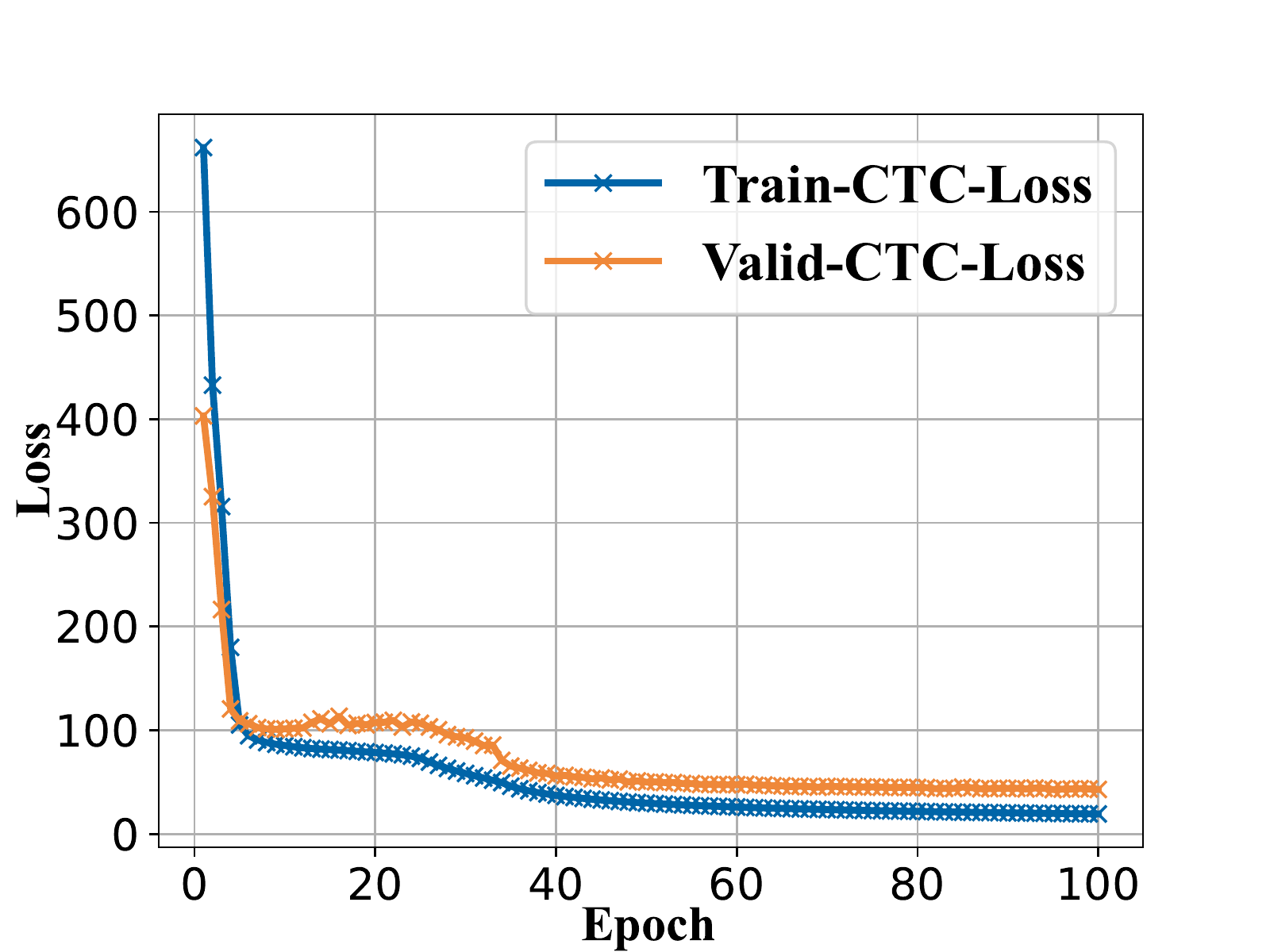}

    \end{minipage}
    \hfill
    \begin{minipage}[t]{.485\linewidth}
      \includegraphics[width=1\linewidth]{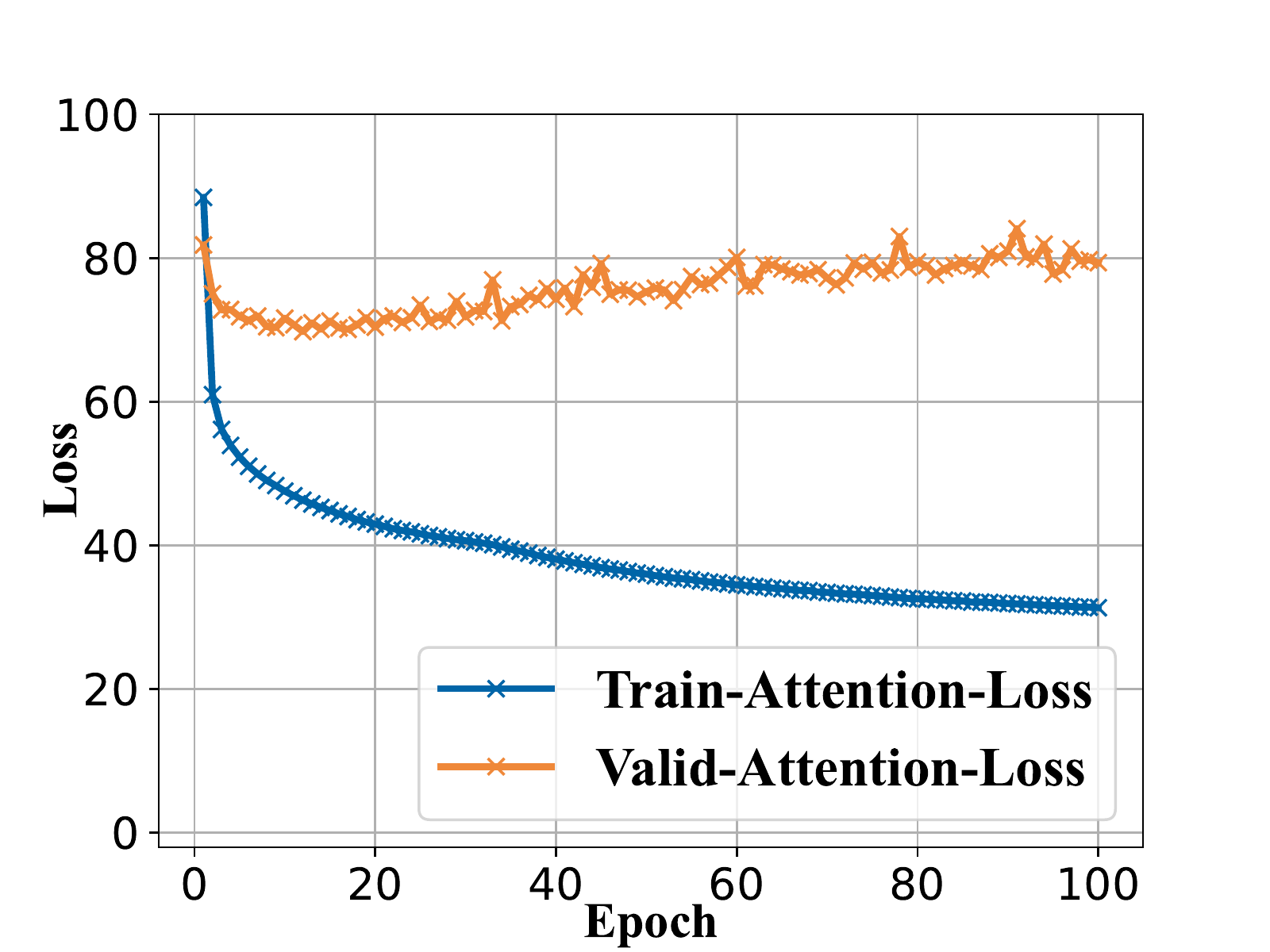}
    \end{minipage}
  \caption{Loss vs training epoch with the criteria of CTC versus attention models.}
  \label{fig:training-loss-curve}
\end{figure}

Table~\ref{tab:multilingual-result} shows us that CTC-Attention joint learning based multilingual ASR models can also recognize code-switching speech, which beyond our previous knowledge~\cite{khassanov2019constrained}, so we are curious about which part of the ASR model helps to do so. 
Figure~\ref{fig:training-loss-curve}
plots the Loss versus epoch for CTC-only criterion on the left and the  Attention-only criterion on the right.  
% Train is multilingual speech and valid is CS speech. 
The ``Train-*" and ``Valid-*"  refer to  multilingual and CS speech respectively.
We find that for CTC-only system, losses for Train and Valid descend following the same tendency, probably thanks to the non-auto-regressive property of CTC. In contrast, Attention-only system shows a diverged losses on validation data instead, which suggests attention has difficulty to perform CS ASR if CS utterances has never been seen during training process.

% \textcolor{red}{
% We find that for CTC-only system, losses for train/valid descend follow the same tendency, benefit by its conditional independence. In contrast, for Attention-only system, the loss for valid increases while the one for train goes down, which tells us that it is hard for Attention to output code-switching sequence if code-switching utterance is never seen by such model.}

%%%%%%%%%%%%%%%%%%%%%%%%%%%%%%%%%%%%% plots the new figure now......
% \begin{figure}[th]
%     \centering
%     \includegraphics[width=1.0\linewidth]{LaTeX/loss.png}
%     \caption{Loss vs Epoch under CTC or Attention training criterion} 
%     \label{fig:training-loss-curve}
% \end{figure}

%%%%%%%%%%%%%%%%%%%%%%%%%%%%%%%%%%%%% plots the new figure now......

% \begin{table}[htp]
%  \centering
% \caption{Recognition Results of Multilingual ASR models conditioning on CTC weight}
%  \label{tab:multilingual-result-ctc}
% \begin{tabular}{lcccccc}
%  \toprule
%  \multirow{2}*{\makecell[c]{CTC weight}} & \multicolumn{2}{c|}{SEAME} & \multicolumn{3}{c}{ASRU} \\
%  & dev\textsubscript{man} & dev\textsubscript{sge} & dev1 & dev2 & test \\
%  \midrule
%  0.0 (Att only) & \\
%  1.0 (CTC only) & 42.3 & 48.4 & 68.3 & 67.6 & 67.1 \\
%  0.3 & 33.0 & 38.4 & 65.2 & 64.7 & 63.9 \\
%  \bottomrule
% \end{tabular}
% \end{table}

\section{Conclusion}~\label{sec:con}
In this paper, we proposed an ILME-based LM fusion for cross-domain CS ASR. We have demonstrated the effectiveness of the proposed method on SEAME  and a so-called ASRU Mandarin-English code-switching  corpora. We also found the proposed IMLE-based LM fusion method is also very effective on the ASR models that are obtained by combining two data sets, in terms of intra-domain CS ASR. More importantly, we have also attempted to train a multilingual ASR system by merging two monolingual data,  to perform CS ASR. With the help of the proposed method, as well as the joint CTC-attention training criterion, the performance improvement is very significant, achieving up to 32.1\% mixed error rate reduction compared with conventional shallow fusion method.
% In this paper, we investigated language model fusion techniques and internal language model estimation methods applying to Code-Switching speech recognition. We proposed two different ILME methods to boost the performance of the State-of-the-art Conformer models and Shallow Fusion methods in Code-Switching Scenario. 
% Consequently, the proposed methods achieved consistently improvements on all the evaluation sets on two different CS domain. Finally, we verified that CTC-only model is capable of recognizing CS speech while it was trained using only mixed monolingual data.
% We also verified that CTC module is capable of recognizing CS speech while it was trained using only mixed monolingual data. 
%The loss suggested that it is hard for Attention to do such job alone.   

\vfill\pagebreak
\clearpage
\bibliographystyle{IEEEtran}

\bibliography{mybib}

\end{document}